# Nanobubble induced formation of quantum emitters in monolayer semiconductors

Gabriella D. Shepard[1], Obafunso Ajayi[2], Xiangzhi Li[1], X.-Y. Zhu[3], James Hone[2], and Stefan Strauf[1*]

[1]Department of Physics & Engineering Physics, Stevens Institute of Technology, Castle Point on the Hudson, Hoboken, New Jersey 07030, United States

[2]Department of Mechanical Engineering, Columbia University, New York, New York 10027, United States

[3]Department of Chemistry, Columbia University, New York, New York 10027, United States

*Address correspondence to: strauf@stevens.edu

## Abstract

The recent discovery of exciton quantum emitters in transition metal dichalcogenides (TMDCs) has triggered renewed interest of localized excitons in low-dimensional systems. Open questions remain about the microscopic origin previously attributed to dopants and/or defects as well as strain potentials. Here we show that the quantum emitters can be deliberately induced by nanobubble formation in $WSe_2$ and $BN/WSe_2$ heterostructures. Correlations of atomic-force microscope and hyperspectral photoluminescence images reveal that the origin of quantum emitters and trion disorder is extrinsic and related to 10 nm tall nanobubbles and 70 nm tall wrinkles, respectively. We further demonstrate that "hot stamping" results in the absence of 0D quantum emitters and trion disorder. The demonstrated technique is useful for advances in nanolasers and deterministic formation of cavity-QED systems in monolayer materials.

**Introduction**

Since their successful isolation from bulk crystals, TMDC monolayers have been studied extensively because they possess desirable intrinsic properties such as a direct band gap [1] and the absence of an inversion center [2], which make these 2D materials promising hosts for photonic [3], optoelectronic [4,5], and valleytronic applications [6–10]. A key advantage compared to traditional compound semiconductors is the strong Coulomb interactions caused by reduced dielectric screening in these monolayer materials resulting in large exciton binding energies approaching 0.6 eV [11–13] as well as trion formation with binding energies up to 20 meV [11], thereby enabling room temperature applications. Of particular interest are recent demonstrations of quantum light emitters in these materials that are characterized by strong single photon antibunching signatures, zero-phonon exciton linewidth down to 100 µeV, and large gyromagnetic ratios [14–18].

Open questions remain about the microscopic origins of these localized excitons in 2D materials in order to control them for devices applications. In general, mobile excitons in semiconductors can either get localized at intrinsic or extrinsic dopant atoms or they can be energetically trapped due to extrinsically induced quantum confinement. In the first case excitons form an immobile effective-mass four-particle complex with the dopant atoms known as impurity-bound excitons as originally predicted by Lampert [19]. In the second scenario excitons localize within low-dimensional materials due to local quantum confinement, for example in 2D by quantum-well width fluctuations [20] or in 0D quantum dots trapping individual excitons that give rise to quantum light emission [21]. The first demonstration of quantum light from impurity bound excitons more than a decade ago could directly pinpoint the microscopic origin to substitutional nitrogen acceptors located on a selenium lattice site [22]. Similarly, recent work on oxygen-doped carbon nanotubes could clearly identify the quantum light emission to exciton localization at intentionally introduced ether-d and epoxide-I solitary-dopant atom configurations that create 130-300 meV deep exciton trap states [23]. For $WSe_2$ and $MoSe_2$ monolayers it is apparent that spectrally broad defect or dopant-activated photoluminescence (PL) bands exist [24,25] that could be responsible for the spectrally sharp localized exciton emission at cryogenic temperatures. It was however also found that the quantum emitters in $WSe_2$ predominantly appear near 2D layer interfaces, edges, and wrinkles[14–18], and can be deliberately induced by tearing the monolayer lattice[23] or by using metal nanogaps underneath the monolayer material[25]. Towards more systematic investigation of their origin, Kumar *et al.* have directly correlated the appearance and localization energy of quantum emitters with the local strain potentials that are induced when monolayers tear over etched holes in the substrate[24]. If exciton quantum emitters are strain-induced in monolayer semiconductors, i.e. are of external origin, it should be possible to deliberately induce quantum emitters with controllable density due to stressor points from the substrate without structurally damaging the TMDC layer.

Here we demonstrate that quantum light emitters with high areal densities can be formed from deliberately introduced nanobubbles in monolayer TMDCs. By correlating optical images, atomic force microscopy, and hyperspectral photoluminescence images, we find structural features of typically 5-10 nm that we classify as nanobubbles leading to pockets containing either individual or in some cases up to about four quantum emitters. Furthermore, we present a hot-stamping technique that effectively removes the quantum emitter occurrence, thereby providing a clean slate for applications such as spatially deterministic placement of quantum light emitters via lithography as well as for optoelectronic applications based on 2D excitons where energetically deeper emission centers would be a detrimental loss channel.

**Results and Discussion**

With the goal to create a large density of spatially separated locations that each trap an individual quantum dot like exciton, we deliberately induced nanobubbles during the dry-stamping process used to transfer 2D layers and assemble heterostructures. As has been known for some time, nanobubbles can be forced into graphene and TMDC layers by applying high pressure to the viscoelastic stamp in the transfer process [26,27]. Once the stamp is released, the sudden strain relaxation leads to formation of bubbles. We refer to this process as "cold stamping" as it is often carried out at ambient temperature and without thermal annealing. In contrast, one can also carry out dry stamping at elevated substrate temperatures and by adding overnight thermal annealing in an ultra-high vacuum chamber in each step, a process referred to in the following as "hot stamping" (see Methods). We will first describe experiments utilizing cold stamping of both $WSe_2$ monolayers and $BN/WSe_2$ heterostructures onto $Si/SiO_2$ substrates.

The optical microscope image in **figure 1a** of the cold-stamped sample studied shows the heterostructure appearing slightly blue (**$WSe_2$/BN**) and the monolayer regions as dark purple areas (**1L**). Judging from the optical image alone, the resulting sample is free of wrinkles nor is there visible surface contamination. To characterize the exciton emission, low-temperature (3.8 K) micro-photoluminescence (µ-PL) spectra were taken at characteristic locations on both the heterostructure and monolayer (**figure 1b**). When exciting with moderately high pump powers, the monolayer spectrum displays two high energy peaks resulting from the quantum-well-like neutral excitons (**2D-$X^0$**) and charged excitons (**2D-$X^T$**), which are separated by approximately 36 meV, corresponding to the 2D-$X^T$ binding energy. Additionally, there is a broad, low energy feature that is a result of the bound excitons (**BX**) [13,25]. By decreasing the excitation power, the weak BX emission becomes the main feature of the PL emission, and spectrally sharp emission lines appear. Similar pump power dependent bound exciton emission signatures have previously been found dominantly at edges, interfaces, and folds, as well as areas of high strain potentials [14–18,28,29]. In contrast, we find here that the spectrally sharp emission lines are distributed in our samples over the entire interior of the 1L region basal plane. While this might be somewhat surprising on a first glance since the optical microscope image looks clean and free from structural defects, we show in the following that they are a result of nanobubble formation.

To provide more insight into the exciton emission properties, we utilized 2D hyperspectral imaging to address each excitonic species (2D-$X^0$, 2D-$X^T$, BX) separately (see Methods). **Figure 2a** shows the scan for the 2D-$X^0$ emission that retraces the optical image and has a rather uniform distribution showing up everywhere where there is $WSe_2$ while the bare BN regions appears dark, as expected from an insulator with a bandgap of 6 eV. The PL scan also reveals a crack in the center of the $WSe_2$/BN region not visible in the optical image that is dominated by the optical contrast from BN. Relative to the 1L region, the 2D-$X^0$ emission appears weaker in the heterostructure region. This effect is however not caused by PL quenching but simply by an underlying 5 nm spectral redshift of all exciton transitions due to the local difference in strain and dielectric screening when transitioning from the 1L to the $WSe_2$/BN region (see Supplementary Data **figure S.1**), effectively shifting the 2D-$X^0$ peak to the point where the 10 nm bandpass filter transition is reduced by half.

In contrast, the 2D-$X^T$ (**figure 2b**) scan is affected by large spatial disorder on the heterostructure region with a grainy intensity structure, while the 1L region is rather uniform. Though not obvious in the optical image, AFM scans reveal that the heterostructure region has a very non-uniform surface structure containing triangular wrinkles up to 70 nm tall at a high density of approximately 3.3 wrinkles per µm$^2$ (see Supplementary Data **figure S.2**). Wrinkles of comparable size have been previously shown via chemical element analysis to contain pockets of hydrocarbons, a result of the "self-cleaning" squeezing mechanism that occurs due to the van der Waals forces between the 2D materials [30,31]. Density functional theory calculations show further that the bending strain of the TMDC membrane can induce local charge accumulations by shifting the Fermi level [32]. It is thus plausible that hydrocarbon contamination and/or the local strain of these wrinkles result effectively in local electrostatic gating that is responsible for the observed disorder of the trion emission intensity in this area.

The four panels **figure 2c-f** show PL images of the BX emission on the sample where each panel is a 10 nm slice of the spectral region. It should be noted that even when there are no distinct sharp emitters, there is always a weak, broad feature present centered around 1.67 eV (742 nm). As a result, in the 740 map, there is an almost omnipresent signal over the sample, stronger on the monolayer than on the heterostructure. As the spectral filtering moves further out from the center of the omnipresent BX emission, more isolated bright PL pockets are visible, as shown in **figures 2d-f**. More importantly, at the longer wavelengths, corresponding to deeper localized excitons below the 2D bandgap, it becomes more likely that these bright pockets will contain only one quantum emitter.

To verify the quantum nature and quantum-dot like confinement of these bright PL spots found by hyperspectral imaging, we analyze the pocket outlined in **figure 2f** in detail. A high-resolution spectrum of the emitter is shown in **figure 3a**. Individual peaks show spectral diffusion limited linewidth down to about 170 µeV. Further, a doublet structure is visible with two peaks of unequal intensity and a zero-field splitting $\Delta_0$ of 760 µeV, which is a result of electron-hole spin-exchange interaction as well as an underlying anisotropic strain [28]. The doublet's behavior in an external magnetic field of up to 9T applied in a Faraday geometry is shown in **figure 3b**. The Zeeman peak splitting fits to the well-known relation

$$\Delta_B = \sqrt{\Delta_0^2 + (\mu_B g B)^2},$$

where $\mu_B$ is the Bohr magneton and $g$ is the exciton g-factor, revealing g=8.6±0.2. This g factor is almost twice larger than the reported values of about 4-5 for the 2D-X and 2D-$X^T$ excitons [33] and in good agreement with recent reports of quantum emitters in TMDCs [14–18,28]. To directly prove the quantum nature of the light emission we have recorded the second order photon correlation function $g^2(\tau)$ under nonresonant excitation (**figure 3c**). We observe strong single photon antibunching from this emitter with $g^2(0)=0.2$. Additionally, the emitter has a photon recovery time of 3.8 ns based on the $g^2(\tau)$ fit rise time, in agreement with values reported for PL lifetimes (1.5-2.5 ns) [14].

After confirming the quantum nature of the emitters, we now determine their origin and investigate why the monolayer is showing strong BX emission towards the interior of the flake. A 2D map of the BX emission in the monolayer region of the flake is acquired at a very low excitation power, ensuring there is no contribution from the 2D excitons in the signal. Additionally, we take an AFM scan of the sample to facilitate a direct comparison between any structural features on the surface of the monolayer and the bright BX PL pockets. **Figure 4** shows an overlay of the two scans, with the transparency of the PL image changing by 20% in each successive image to aid in finding correlations between the two scans. Similarly, the boundaries of the monolayer flake have also been highlighted in black for clarity. Most notably, the AFM shows that the sample is structured, as expected from the cold-stamping approach. There are multiple structures visible on the monolayer, including several 3-6 µm long wrinkles located near the BN interface, four circular 60-70 nm tall structures, and about 25 nanobubbles. Of the visible features, it is the smaller nanobubbles of 4-10 nm height that consistently match up to the PL pockets, where 19 out of 20 PL pockets contain at least one nanobubble. We note that not all features correlate to optical emission, in particular not the four circular features that are with 60-70 nm height significantly taller than the nanobubbles, possibly because they are on top instead of underneath the TMDC material, which cannot be distinguished in the AFM scans.

To further quantify the correlation of nanobubble location with occurrence of BX quantum emitters, we studied the exciton emission spectra of the PL pockets highlighted in **figure 4a** and correlated it with the structural feature from the AFM. The near solitary quantum emitter spectrum from P1 shown in **figure 5a** correlates with an individual nanobubbles of 10 nm height and 900 nm lateral extent as shown **figure 5b**. Regions that appear flat and thus free of bubbles or wrinkles in the AFM scans, as indicated by the red trace in **figure 5b**, show no distinct sharp BX lines. Likewise, the spectrum for P2 in **figure 5c** contains about eight sharp lines, or four quantum emitter doublets, that correlate with a chain of 3-4 nanobubbles reaching up to a height of 6 nm, shown in **figure 5d**. Additional cases for P3 and P4 are shown in the supplementary data **figure S3.**

Having established that deliberately induced nanobubbles via cold stamping are responsible for localized quantum emitters in TMDCs, as well as that structural folds are responsible for 2D-$X^T$ disorder, it is of interest to have a method for controlling the density of these structural features, ideally down to zero for the production of pristine flakes free of 0D-BX emission. To this end, we have utilized a hot stamping procedure that employs thermal annealing at 315 C. The higher temperature allows pockets of residue between layer interfaces to flow and spread out, effectively leaving regions of atomically clean interfaces[30]. Therefore, in an attempt to make a surface free from the nanobubbles or wrinkles we have fabricated a BN/WSe$_2$ heterostructure by hot stamping and AFM imaged it before and after annealing to demonstrate that structural disorder can be removed post dry-stamping (see Supplementary Data **figure S4**).

For a similar sample fabricated by hot stamping from the same WSe$_2$ host crystal as the data presented for cold stamping, we show in **figure 6a-c** scans complementary to the hyperspectral images from **figure 2**. While the 2D-X map shows high uniformity, as expected, the emission from 2D-$X^T$ now lacks the strong disorder seen in the cold-stamped sample (**figure 2b**), demonstrating that it can be removed by the hot stamping technique. Similarly, hyperspectral imaging focusing on BX emission does not show any bright PL pockets and only a very dim background emission that is distributed uniformly over the heterostructure (**figure 2c**). Like the cold-stamped sample, there is a monolayer and heterostructure area, and AFM scans over the heterostructure show the sample lacks the large 70 nm wrinkles and instead has a smooth surface with only small 1-2 nm fluctuations. Similarly, the monolayer piece of the hot-stamped flake shows almost a surface roughness of less than 1 nm. **Figure 6d-e** show the comparison between the typical features found on the cold stamped flake compared to the hot-stamped sample. A typical PL spectrum from the hot-stamped sample is compared to **P2** in **figure 6f**. Clearly, no sharp quantum emitters are visible and instead a broad and rather weak BX emission is present.

Finally, we provide a qualitative model for the microscopic origin of the localized exciton emission in WSe$_2$. On one hand, we have shown a clear correlation between the presence of nanobubbles and spectrally sharp quantum emitters as well as the absence of these 0D quantum emitters in hot stamped and structurally smooth samples. This clearly indicates that 0D quantum emitters in WSe$_2$ can be extrinsically induced. On the other hand, it is also apparent that a residual broad background emission remains in the absence of nanobubbles or structural folds, as is evident from the hot stamping results in the red trace in **figure 6f**. It is thus plausible to assume that a residual broad emission band centered around 1.675 eV (740 nm) has its origins in native defects, predominantly Se vacancies as previously reported [25]. The considerable density of vacancies that naturally occurs within an optical probe area of about 1 μm gives rise to an omnipresent impurity-bound exciton emission signature in $WSe_2$ that is characterized with localization energies of 70 meV below the 2D neutral exciton emission (1.745 eV). As can be seen in the red trace in **figure 6f**, several sharper lines ride on top of this broader background, but do not go beyond the Gaussian envelope of the sub-ensemble emission, indicating the onset of the break up into individual impurity bound excitons, similar to the original studies on acceptor-bound quantum light emission with intermediate dopant density [22]. In stark contrast, the presence of external stressor centers, such as the nanobubbles that we induced in high density over the interior flake, leads to strain-induced localization of excitons with localization energies up to 165 meV (see Supplementary Data **figure S2a**). These extrinsically induced quantum emitters have not only a much stronger oscillator strength, i.e. brighter intensity, but appear predominantly on the low-energy side of the Gaussian envelope of the impurity bound exciton emission (see blue trace in **figure 6f**), supporting the picture of a strain-induced quantum confinement mechanism as suggested previously [28]. By the nature of the lateral extent of the strain induced confinement potential of several hundred nm, individual nanobubbles can host more than one exciton quantum emitter, but typically we find not more than four quantum emitters per nanobubble. In that sense, one can draw an analogy to InAs self-assembled quantum dots that are known to localize excitons out of their adjacent 2D wetting layer due to a local three-dimensional quantum confinement and likewise occur always energetically lower than the emission from their adjacent wetting layer [34].

**In summary**, we have shown correlations between optical images, atomic-force microscopy, and hyperspectral photoluminescence images, which reveal the origin of the quantum light emission to be extrinsic to the monolayer material. The quantum dot-like emitters appear to be a direct result of nanobubble formation that can be deliberately created at high density during fabrication in a cold-stamping process. The number of spectral doublets in an optical hot spot seems to correlate with the number of nanobubbles located within these PL regions. These quantum emitters are characterized by pronounced single photon antibunching with $g^2(\tau)=0.2$, narrow spectral linewidth down to 117 μeV, and large exciton g-factors of about 8.6. In addition, we have demonstrated that the trion emission can be largely disordered in the $WSe_2$/BN heterostructure and is correlated with the presence of a high density of 70 nm tall wrinkles that can form between heterostructure layers. We further have demonstrated that disorder in trions and the occurrence of quantum emitters can be fully removed by a hot stamping procedure, producing TMDC material that can act as a clean platform for future deterministic fabrication of cavity-QED systems via lithography as well as for optoelectronic applications based on 2D excitons where energetically deeper emission centers would be a detrimental loss channel. The demonstrated controlled formation of 0D excitons at high density within monolayer material and without structurally damaging them could also open up new routes to create efficient quantum dot nanolasers from 2D materials when integrated into cavities [34].

Note that after this work was completed we became aware of two related papers on arxive [35,36] reporting externally induced formation of exciton quantum emitters in TMDC materials based on patterning nanopillars on a substrate that act as local stressor centers. While the nanopillar lithography approach offers spatially deterministic placement of individual quantum emitters, it requires higher aspect ratios up to 0.15 which can lead to detrimental structural piercing of the monolayer material. In contrast, our demonstrated nanobubble approach does not require lithography and creates quantum emitters at an order of magnitude lower aspect ratio leaving the lattice intact, while the demonstrated hot stamping technique is of advantage for both approaches.

## Materials and Methods

*Sample preparation and stamping procedure*

WSe$_2$ from 2D semiconductors was mechanically exfoliated onto a viscoelastic stamp (Gel-Film®) using Nitto Denko tape. Thin layers were identified by their optical contrast using an optical microscope. The same was done for BN from HQ Graphene. The heterostructure was stamped onto a standard Si wafer with 90 nm oxide, and the surface was cleaned and made hydrophilic to improve adhesion between the flake and the substrate. To achieve this, the SiO$_2$ substrate was submerged in 30% KOH solution for 20 minutes and rinsed in DI water for 3 minutes. Stamping was done immediately after surface preparations were completed. Cold-stamping was done at room-temperature, with the BN stamped down first followed immediately by stamping of the WSe$_2$ flake. In contrast, hot-stamping included thermal annealing steps between each stamping process in a homebuilt ultra-high vacuum chamber with button heater (HeatWaveLabs). First, the BN was stamped onto the clean SiO$_2$ surface, then annealed at 315 C for approximately 12 hours. Afterwards, WSe$_2$ is stamped over the BN and annealed in the chamber again at 315 C for another 12 hours before optical measurements have been taken.

*Optical measurements*

Micro-photoluminescence (µ-PL) and magnetic field measurements were taken inside a closed-cycle He cryostat with a 3.8 K base temperature and a superconducting magnet that can apply a field up to 9 T (attodry1100). Samples were excited with a laser diode operating at 532 nm in continuous wave mode. An Abbe limited laser spot size of about 500 nm was achieved using a cryogenic microscope objective with numerical aperture of 0.82. The relative position between sample and laser spot was adjusted with cryogenic piezo-electric xyz-stepper (attocube) while 2D hyperspectral scan images were recorded with a cryogenic 2D-piezo scanner (attocube) typically run at 100 µm/s and by filtering through 10 nm bandpass filters before entering a multimode fiber attached to a single photon counting avalanche photodiode (silicon APD). Regular PL spectra from the sample were collected in the same multimode fiber, dispersed using a 0.75 m focal length spectrometer with either a 300 or 1200 groove grating, and imaged by a liquid nitrogen cooled silicon CCD camera. Laser stray light was rejected using a 532 nm RazorEdge ultrasteep long-pass edge filter.

*Photon correlation measurements*

The second-order photon correlation function $g^2(\tau)$ was recorded by time-correlated single photon counting (TCSPC) sending the PL emission through narrow bandpass filters and onto a Hanbury-Brown and Twiss setup consisting of a fiber-coupled 50/50 beam splitter connected to two silicon APDs. A linear polarizer was used in the collection path for photon antibunching measurements. Coincidence counts were time stamped and analyzed with a high-resolution timing module (SensL).

*Atomic force microscope (AFM) imaging*

The AFM measurements were obtained using a Bruker Dimension FastScan AFM in noncontact mode at a scan rate of 1 Hz with a FastScan-B AFM tip that has a nominal spring constant of 1.8 N/m. Height profiles were extracted from the AFM images using Gwyddion open source software.

## Acknowledgements
We like to thank Milan Begliarbekov for supporting the AFM measurements at the City University of New York Advanced Science Research Center (ASRC) nanofabrication facility and for machining the UHV heater-assembly. S.S. and J.H. acknowledge financial support by the National Science Foundation (NSF) under awards DMR-1506711. S.S. acknowledges financial support for the attodry1100 system under NSF award ECCS-MRI-1531237. O.A., X.Y.Z., and J.H. acknowledge financial support from the NSF MRSEC program through Columbia in the Center for Precision Assembly of Superstratic and Superatomic Solids (DMR-1420634).

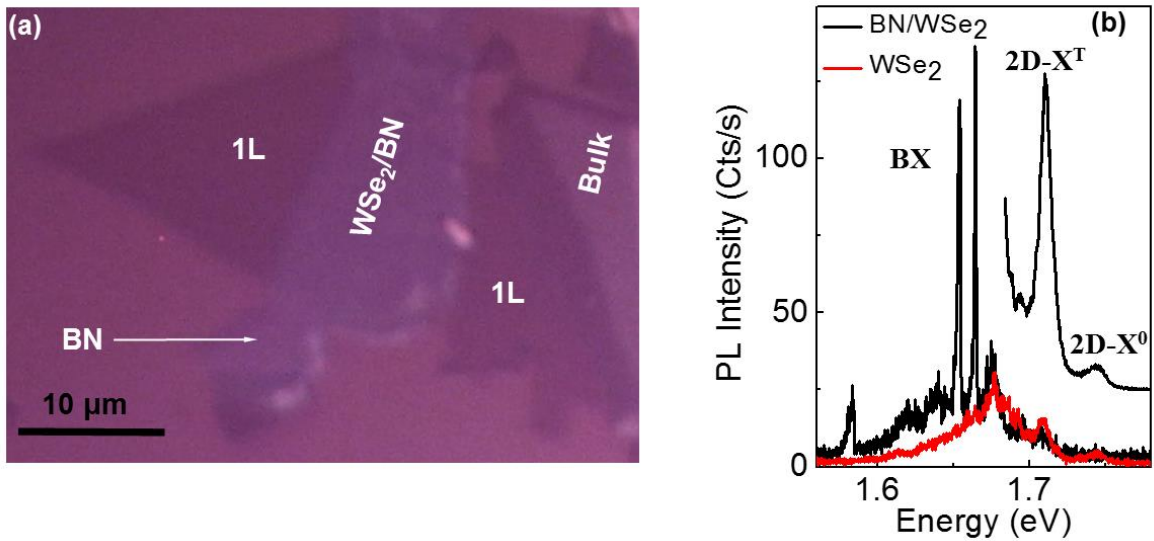

**Figure 1. Optical emission from cold-stamped monolayer WSe$_2$ and BN/WSe$_2$ heterostructure.** (a) Optical image of a cold-stamped WSe$_2$ monolayer (**1L**) and BN/WSe$_2$ heterostructure (**WSe$_2$/BN**) on a 90-nm SiO$_2$ substrate. (b) Photoluminescence spectra recorded at 3.8 K under nonresonant excitation (532 nm) of monolayer WSe$_2$ and a BN/WSe$_2$ heterostructure. The bound excitonic emission (**BX**) occurs below 1.7 eV and dominates the signal at low pump powers (250 nW). The spectrum shown as an inset on the upper right area was recorded at five-fold higher pump powers of 1.2 µW in order to clearly show the neutral (**2D-X$^0$**) and charged (**2D-X$^T$**) exciton emission that dominates at higher pump powers.

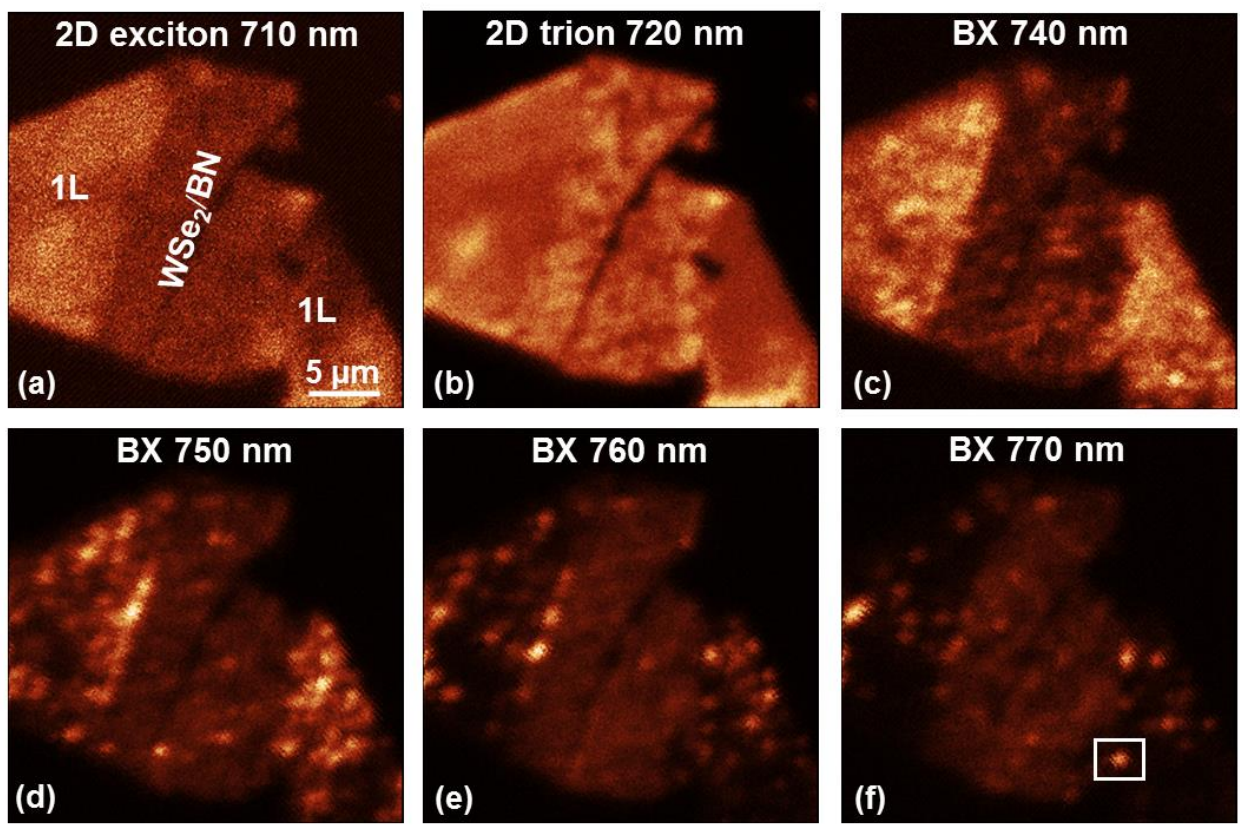

**Figure 2. Hyperspectral imaging of the monolayer and heterostructure.** 2D maps of the PL emission taken over the spectral range 710-770 nm. Each map is taken by exciting the sample nonresonantly (532 nm), with the resulting emission filtered using a 10-nm bandpass centered on the indicated wavelength. The spatial scans over the **2D-X$^0$** and **2D-X$^-$** emission (a-b) are taken at an excitation power of 23.5 nW, while the **BX** scans (c-f) are taken at lower power of 2.5 nW. The white box in (f) indicates a localized **0D-X** line that is studied in more detail.

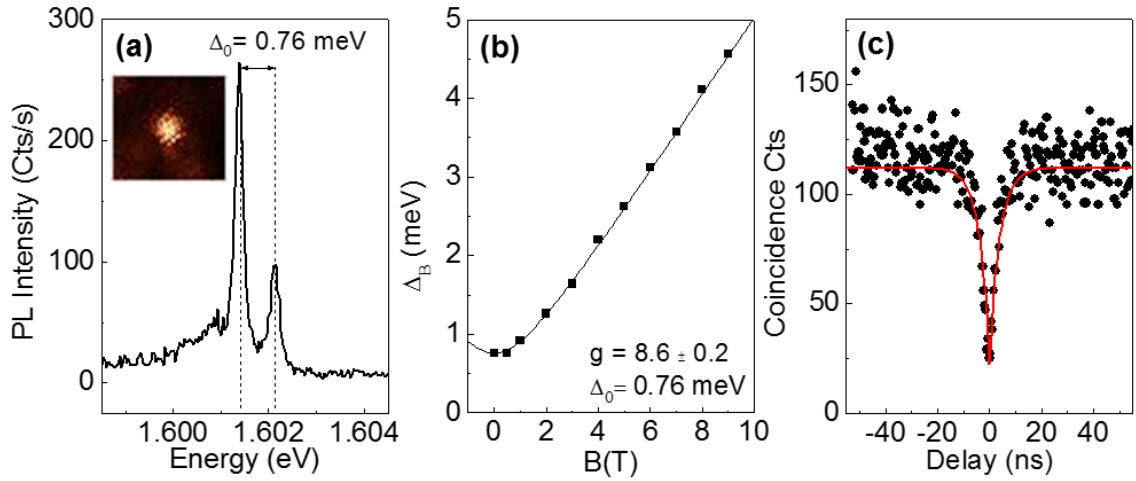

**Figure 3. Magnetic field dependence and antibunching for a typical 0D-X line.** (a) High resolution spectra of **0D-X** line from the location highlighted in **figure 2**. The inset is the 2D spatial PL scan of this emitter. The two peaks of the doublet are the zero-field fine structure splitting components that have unequal intensities. (b) Magnetic field dependence of the Zeeman splitting up to B = 9T in Faraday geometry. (c) Second order correlation function $g^{(2)}(\tau)$ of the **0D-X** line under nonresonant CW excitation showing pronounced photon antibunching with $g^{(2)}(0) = 0.2$ and a long photon recovery time of 3.8 ns. Out to 1 µs delay time there is no apparent bunching behavior visible indicating that the quantum emitter is not affected by blinking.

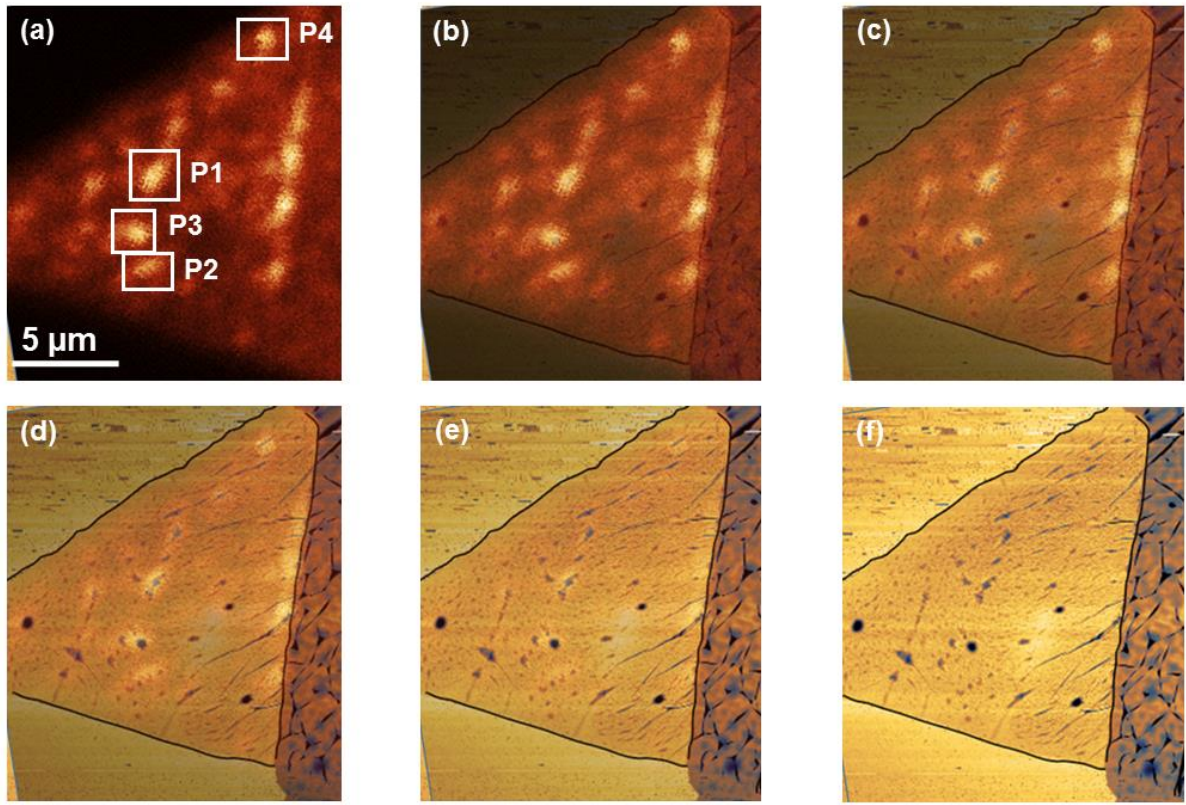

**Figure 4. Correlation between the $X^B$ PL map and AFM scan for the WSe$_2$ monolayer.** (a) Overview PL and (b) AFM scans of the WSe$_2$ monolayer region. The PL signal was not spectrally filtered for a specific wavelength range but contains only **BX** emission due to the low excitation power (0.7 nW). The two images are overlaid to highlight any correlations between structural features on the sample surface shown in the AFM to the bright PL emission pockets. Panels (b)-(e) change transparency by 20% to fully shift between the two images. The edges of the monolayer, as well as the heterostructure interface, are artificially darkened to aid in overlaying the images, and the color of the AFM has been inverted such that tall features appear dark. P1-P4 indicate bright areas of individual features that are further investigated.

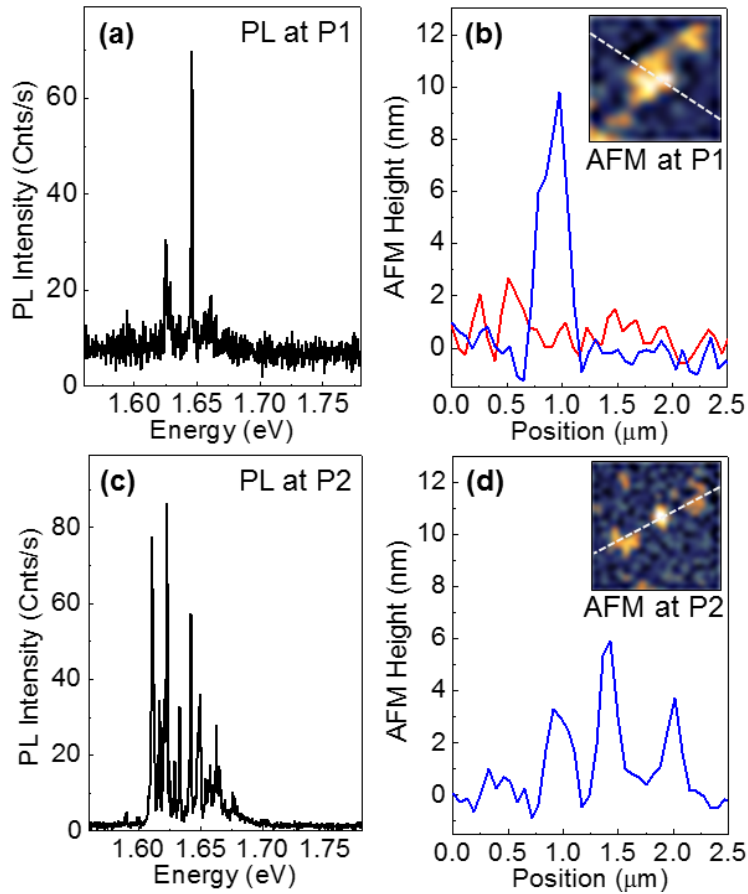

**Figure 5. Emission spectra and AFM height profiles found within bright PL pockets.** (a) Spectra from within bright PL spot P1 indicated in **Figure 4** and (b) the AFM profile for a corresponding nanobubble (blue line). For comparison, the profile from a flat region of the flake is included (red line). (c) and (d) show the spectra and AFM from P2. The insets are 1.5 µm x 1.5 µm AFM scans of the nanobubbles. The dashed line indicates the line cut for each height profile

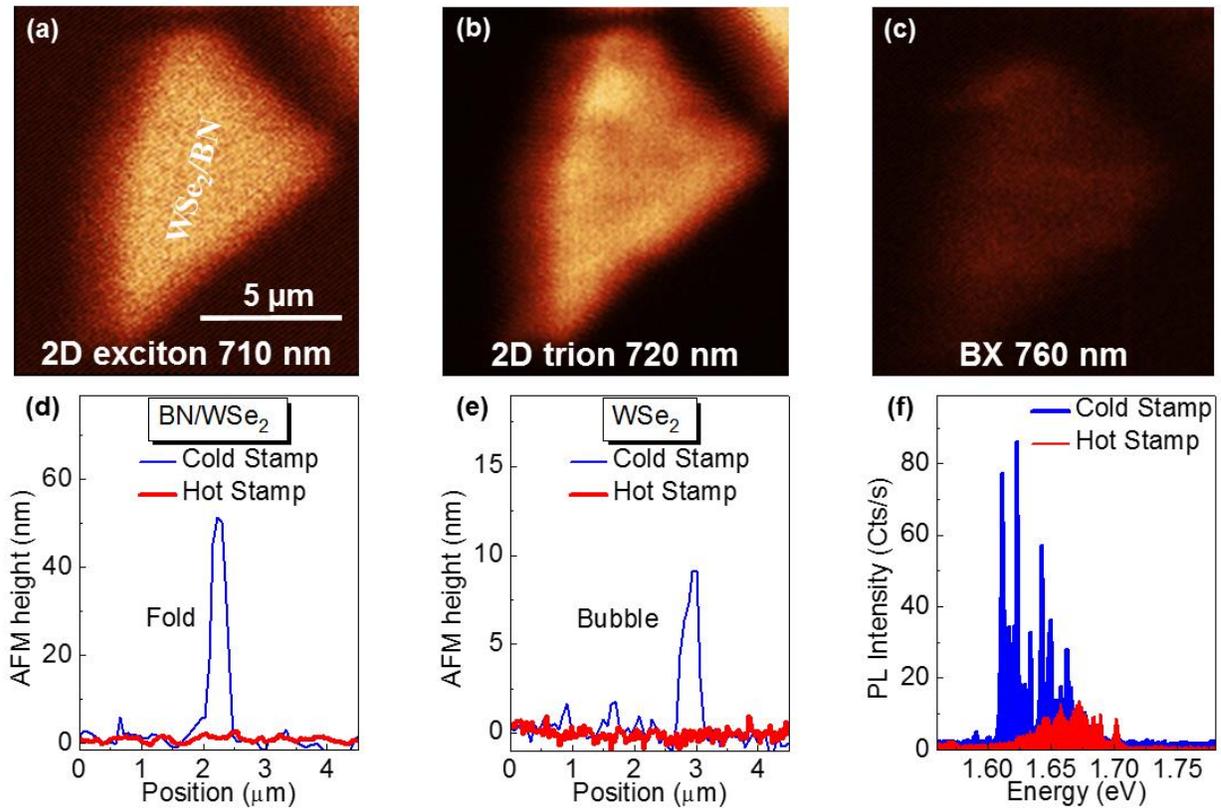

**Figure 6. Impact of hot stamping on PL emission and structural features.** (a)-(c) Hyperspectral imaging of a hot-stamped BN/WSe$_2$ heterostructure. The 2D maps are taken with 10-nm bandpass filters centered over the **2D-X$^0$**, **2D-X$^T$**, and **BX-760 nm** emission. (d) The large 50-70 nm high folds found on the heterostructure in cold-stamping are absent in the hot-stamped heterostructure. (e) The nanobubbles responsible for the **BX** emission found in the cold-stamped WSe$_2$ monolayer are not present in a hot-stamped monolayer. Surface roughness on the hot-stamped monolayer fluctuates by 1 nm or less. (f) Photoluminescence spectra recorded at 3.8 K under nonresonant excitation (532 nm) for the cold and hot stamped flakes. The broad **BX** emission does not structure up into sharp lines on the hot-stamped flake. The hot-stamped PL was taken at an excitation power of 5 nW and the cold-stamped PL at 2.5 nW.